# Realization of all-optical higher-order spatial differentiators based on cascaded operations

Yichang Shou,[1] Yan Wang,[1] Lili Miao,[2] Shizhen Chen,[2] and Hailu Luo[1,*]

[1]*Laboratory for Spin Photonics, School of Physics and Electronics, Hunan University, Changsha 410082, China*

[2]*Key Laboratory for Micro-/Nano-Optoelectronic Devices of Ministry of Education, School of Physics and Electronics, Hunan University, Changsha 410082, China*

*hailuluo@hnu.edu.cn

**ABSTRACT**

Cascaded operations play an important role in traditional electronic computing systems for the realization of advanced strategies. Here, we introduce the idea of cascaded operations into all-optical spatial analog computing. The single function of the first-order operation is difficult to meet the requirements of practical applications in image recognition. The all-optical second-order spatial differentiators are implemented by cascading two first-order differential operation units, and the image edge detection of amplitude and phase objects are demonstrated. Our scheme provides a possible pathway towards the development of compact multifunctional differentiators and advanced optical analog computing networks.

**INTRODUCTION**

Optical analog computing has attracted extensive attention due to its high speed, low power consumption and multi-channel parallel computing capability[1-3]. In recent years, different functions on optical analog computing have been developed such as differentiation[4], solving integration[5] and differential equations[6]. In particular, optical differentiators based on metamaterials and metasurfaces have been extensively reported[7-13]. In addition, the optical differentiation can also be realized with conventional optical interface based on photonic spin Hall effect[14,15] and Brewster effect[16,17]. Differential operations have important applications in all-optical image processing[18], biomedical imaging[19-23], and quantum imaging[24,25]. These differentiators are generally implemented in first-order operations. However, higher-order differentiation operations are required in practical image recognition[26]. In electronic systems, operational amplifiers based on cascaded operations can perform a variety of higher-order operations[27]. By considering the introduction of cascaded operations into optical system, it is expected to realize a multifunctional optical analog computing.

In this paper, we propose a simple method to implement all-optical higher-order spatial differentiators based on cascaded operations. An all-optical second-order spatial differentiator is synthesized by cascading two first-order differentiation units, and the first-order differentiator is composed of a 4*f* system based on the metasurface. Here, we verify the operation function of the second-order differentiator using Gaussian beam input to this cascaded system. Then the image edge detection of amplitude and phase objects are experimentally demonstrated. This scheme allows cascading multiple operation units to synthesize optical analog operators of arbitrary higher-order, which has potential applications in image processing and feature extraction. It also provides a feasible path for the development of compact multifunctional differentiators and advanced optical analog computing networks.

**THEORETICAL ANALYSIS**

A schematic diagram of the all-optical higher-order spatial differentiators based on cascaded operations are shown in Fig. 1. When the incident light is a Gaussian beam, its input electric field is

$$E_{in}(x,y) = \frac{\sqrt{2}}{\sqrt{\pi}w_0} exp\left[-\frac{x^2+y^2}{w_0^2}\right], \quad (1)$$

where $w_0$ represents the beam waist.

Here, the second-order differentiator is formed by cascading two differential units, each performing a first-order differentiation operation in the *x*-direction [Fig. 1(a)]. After the Gaussian beam passes through the first differential unit, the first-order output electric field can be expressed as

$$E_{out}^I(x,y) \propto \frac{\partial E_{in}(x,y)}{\partial x}. \quad (2)$$

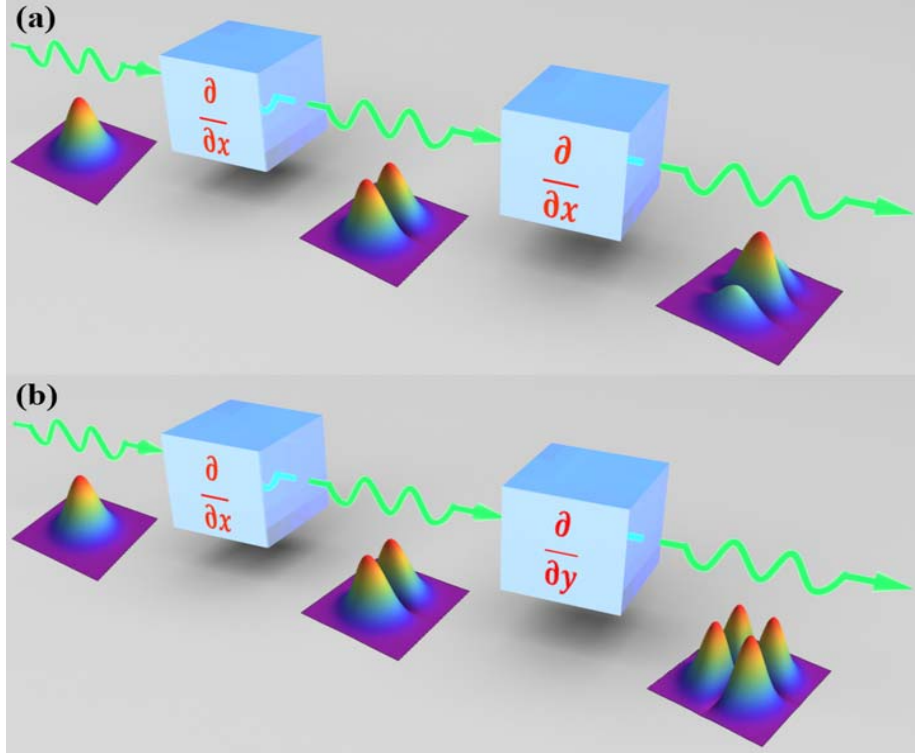

**FIG. 1.** Schematic diagram of a higher-order spatial differentiator based on cascaded operations. (a) Cascaded two first-order differentiators in the x-direction to achieve second-order differentiation $\partial^2/\partial x^2$. The three-dimensional image represents the electric field distribution of the Gaussian beam through this cascaded differentiator. (b) Cascaded two first-order differentiators in different directions to realize second-order differentiator $\partial^2/\partial x \partial y$.

The output of the first differential unit is used as the input of the second unit. After passing through the second unit, the electric field can be expressed as

$$E_{out}^{II}(x,y) \propto \frac{\partial E_{out}^{I}(x,y)}{\partial x}. \tag{3}$$

Substituting Eq. (2) into Eq. (3), the final output electric field corresponding to the second-order differentiation of the initial electric field can be written as

$$E_{out}^{II}(x,y) \propto \frac{\partial^2 E_{in}(x,y)}{\partial x^2}. \tag{4}$$

As shown in Fig. 1(b), the second unit is differentiated for the *y*-direction. The beam accumulation performs two differential operations in different directions, and the second order differential expression is as follows

$$E_{out}^{II}(x,y) \propto \frac{\partial^2 E_{in}(x,y)}{\partial x \partial y}. \tag{5}$$

After cascading *m* first-order differential units in *x*-direction and *n* units in *y*-direction,

the higher-order differential can be obtained as

$$E_{out} \propto \frac{\partial^{m+n}E_{in}(x,y)}{\partial x^m \partial y^n}. \tag{6}$$

The cascaded operations can also be extended to various analog computing units, such as integral units and equation solving units, to realize a multifunctional integrated optical analog computing networks.

To verify the feasibility of implementing higher-order differential operations by cascaded differential units, we design the experimental scheme as shown in Fig. 2(a). The laser generates an input Gaussian beam of $\lambda = 532nm$, and two blue boxes indicate the first-order differential operation unit. The beam passes through two cascaded operation units, and the final second-order differential output intensity distribution is detected on a charge coupled device (CCD). The detection object is placed in the front focal plane of the first lens, and the intensity and phase targets are shown in the inset. The internal structure of the differential unit is shown in Fig. 2(b). Two lenses L1 and L2 with a focal length of $175mm$ form an optical 4f system. The Pancharatnam-Berry (PB) phase metasurface is placed on the Fourier plane of L1 to perform first-order differential operations. The input linear polarization state is prepared by a Glan laser polarizer.

The two circularly polarized components the linear polarized light pass through the PB phase metasurface, when the handedness reversal and spatially uniformly varying PB phase $\Phi_G = -2\sigma_{\pm}\phi(x,y)$ is obtained, where $\sigma_+ = 1$ and $\sigma_- = -1$ $\phi(x,y) = \pi x/\Lambda$ denote left- and right-handed circular polarization, respectively. represents the optical axis direction of the metasurface, and the local optical slow-axis distribution obtained by the cross-polarization method is shown in the Fig. 2(b) inset. $\Lambda = 8nm$ is the period of the etched two-dimensional artificial nanostructure. After a propagation distance $z$, the transverse displacement $\Delta x = z\Delta k_x/k = -\sigma_{\pm}\lambda z/\Lambda$ in real space can be converted by the displacement in momentum space[26], and $\lambda$ is the working wavelength. Here, $\Delta k_x = \partial \Phi_G/\partial x = -2\sigma_{\pm}\pi/\Lambda$ is the spin-dependent momentum displacement arising from the PB phase. The displacement $\Delta y$ can be achieved by rotating the metasurface, i.e., first-order differentiation in the y-direction.

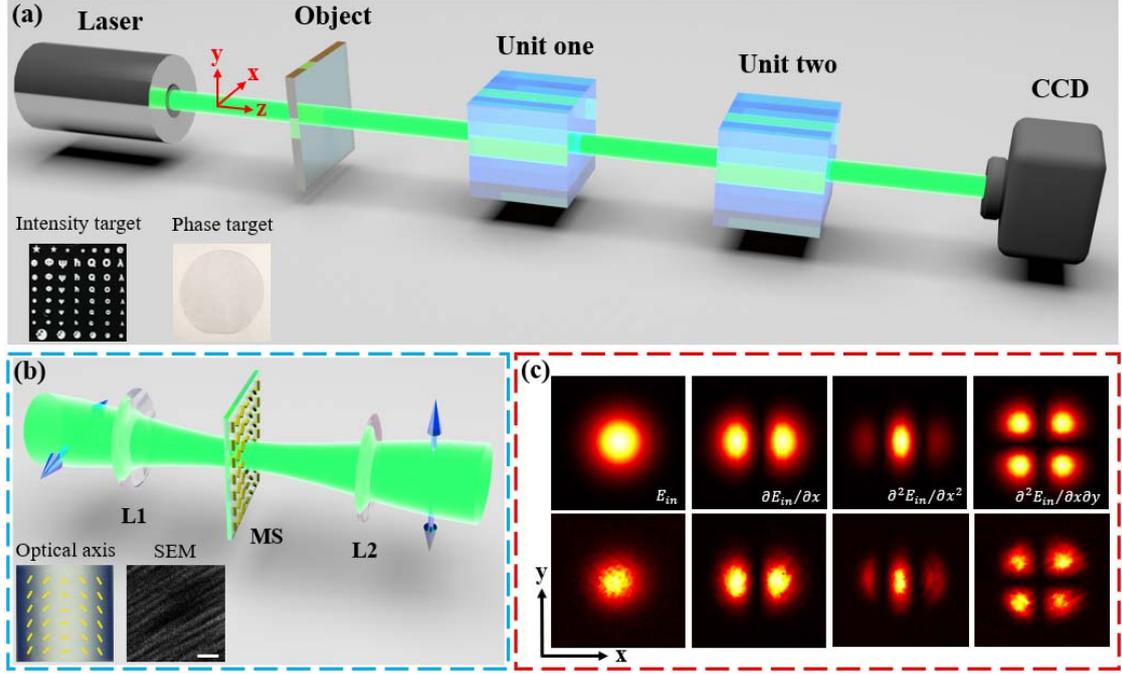

**FIG. 2.** All-optical second-order spatial differentiators based on cascaded operations. (a) Experimental optical path diagram. The incident Gaussian beam is $\lambda = 532 nm$, the object is placed in the front focal plane of the lens, and CCD receives the light field in the back focal plane of the lens. Intensity and phase targets are shown in the inset. (b) The internal structure of the differential unit. The focal lengths of L1 and L2 are $175 mm$; MS, metasurface. Insets are the map of local optical slow axis distribution and scanning electron microscopy (SEM) image of the metasurface nanostructure (scalebar, $300 nm$), respectively. (c) Theoretical (first row) and experiment (second row) results of light intensity distribution for input, first- and second-order differentiations, respectively.

We consider a linear polarized light along the *x*-direction, whose electric field can be written in the circular polarization basis as[28]

$$E_{in}^{1}(x,y) = E_{in}(x,y)\begin{bmatrix}1\\i\end{bmatrix} + E_{in}(x,y)\begin{bmatrix}1\\-i\end{bmatrix}. \qquad (7)$$

When the linearly polarized light is incident on the first differential unit, the wave vector space component is converted to real space by the Fourier inversion, and the output electric field can be expressed as

$$E_{out}^{I}(x,y) = E_{in}[(x+\Delta x, y)]\begin{bmatrix}1\\i\end{bmatrix} + E_{in}[(x-\Delta x, y)]\begin{bmatrix}1\\-i\end{bmatrix}. \qquad (8)$$

An orthogonal polarizer is used behind the metasurface to allow only the *x*-polarized

component to pass. The output electric field distribution can be rewritten as

$$E^I_{out}(x,y) = i[E_{in}(x+\Delta x, y) - E_{in}(x-\Delta x, y)]e_x. \quad (9)$$

When $\Delta x$ is far less than the field distribution, the first-order output electric field can be approximated as

$$E^I_{out}(x,y) \simeq 2\Delta x \frac{\partial E_{in}(x,y)}{\partial x} e_x. \quad (10)$$

The output result of the first-order differentiation can be used again as the input to the second cascaded differentiation unit. If the polarizer after the second metasurface still allows only the x-component to pass, the output electric field is expressed as

$$E^{II}_{out}(x,y) = i[E^I_{out}(x+\Delta x, y) - E^I_{out}(x-\Delta x, y)]e_x. \quad (11)$$

Similarly, when $\Delta x$ is far less than the field distribution, the second-order output electric field can be approximated as

$$E^{II}_{out}(x,y) \simeq 2\Delta x \frac{\partial E^I_{out}(x,y)}{\partial x} e_x. \quad (12)$$

Then the final second-order output electric field is related to the input electric field $E_{in}(x,y)$

$$E^{II}_{out}(x,y) \simeq 4\Delta x^2 \frac{\partial^2 E_{in}(x,y)}{\partial x^2} e_x. \quad (13)$$

If rotating the second metasurface produces a longitudinal displacement $\Delta y$, the relationship between the second-order output electric field and the input electric field is rewritten as

$$E^{II}_{out}(x,y) \simeq 4\Delta x \Delta y \frac{\partial^2 E_{in}(x,y)}{\partial x \partial y} e_y. \quad (14)$$

It should be noted that the above approximation is only hold when $\Delta y$ is much smaller than the input field distribution[4].

The dielectric metasurfaces are formed by femtosecond laser etching of spatially varying nanogrooves in fused silica glass[28]. The nanogroove gratings are distributed about below the glass surface and etched in a square of size 4mm. The glass ($SiO_2$) decomposes into porous glasses $SiO_{2(1-x)}$ and $xO_2$ under direct strong laser light, and its refractive index depends on the intensity of the irradiated laser. The phase delay is $\zeta = 2\pi(n_e - n_o)\hbar/\lambda$, where $\hbar$ is the writing depth. $(n_e - n_o)$ is the induced birefringence, which has a value around $-(3 \pm 1) \times 10^{-3}$. The ordinary and

extraordinary effective refractive indices are given as

$$n_o = \sqrt{Fn_1^2 + (1-F)n_2^2}, \quad n_e = \sqrt{\frac{n_1^2 n_2^2}{Fn_2^2 + (1-F)n_1^2}}. \tag{15}$$

Here, $F$ is the filling factor varying in the range 0.1−0.2, $n_1$ and $n_2$ denote the refractive indices of the two media used to prepare the metasurfaces nanostructures, respectively. Subsequently, the nanostructure image was acquired by SEM [Inset of Fig. 2(b)].

The output optical field of the Gaussian beam through the cascaded second-order differential operation system is shown in Fig. 2(c). The first row are the theoretical calculations of $E_{in}$, $\partial E_{in}/\partial x$, $\partial^2 E_{in}/\partial x^2$ and $\partial^2 E_{in}/\partial x \partial y$ respectively, and the input and output electric field distributions are calculated according to $E(x,y) = \sqrt{I(x,y)}$. Experimentally, the output intensity distribution $I_{out}(x,y)$ is acquired by the CCD. The second row provide the corresponding experimental data, which are in good agreement with the theoretical calculations.

**RESULTS**

To test the all-optical image processing capability of the second-order spatial differentiator, three different intensity patterns are selected as the original input images into this cascaded system [Figs. 3(a1)-3(c1)]. The CCD is placed at the back focus of each differentiator unit lens to obtain the sharpest experimental results. After the first differential unit performs the $\partial E_{in}/\partial x$ operation, the one-dimensional edge detection images are shown in Figs. 3(a2)-3(c2). The acquired edge images are used as the input signal to the second differentiation unit, which performs another first-order differentiation in the *x*-direction, i.e., the $\partial^2 E_{in}/\partial x^2$ operation. The images present adjacent double edge detection effect [Figs. 3(a3)-3(c3)]. The images obtained by changing the second unit differential direction for the $\partial^2 E_{in}/\partial x \partial y$ operation is shown in Figs. 3(a4)-3(c4), and the undetected edge part is consistent with the spot distribution of the second-order differential operation.

In practical image processing, the processed objects often contain intensity and phase information. For a pure phase object with the phase distribution of $\varphi(x, y)$, the electric field can be obtained as $E_{out}(x, y) = exp[i\varphi(x, y)] e_x$, and the light intensity distribution detected on the CCD is $I_{out}(x, y) = |exp[i\varphi(x, y)]|^2$. The bright-field images are hidden in the laser background and pattern shape are barely visible [Figs. 4(a1)-4(c1)]. After passing the first differentiation unit, the output electric field was rewritten as $E_{out}(x, y) \propto \Delta x \partial \varphi(x, y)/\partial x \, exp[i\varphi(x, y)] e_x$ and the corresponding output light intensity is $I_{out}(x, y) \propto |\Delta x \partial \varphi(x, y)/\partial x|^2$. The first-order differentiation in the *x*-direction converts the phase gradient into the intensity contrast images and extract the edge information of the phase objects [Figs. 4(a2)-4(c2)].

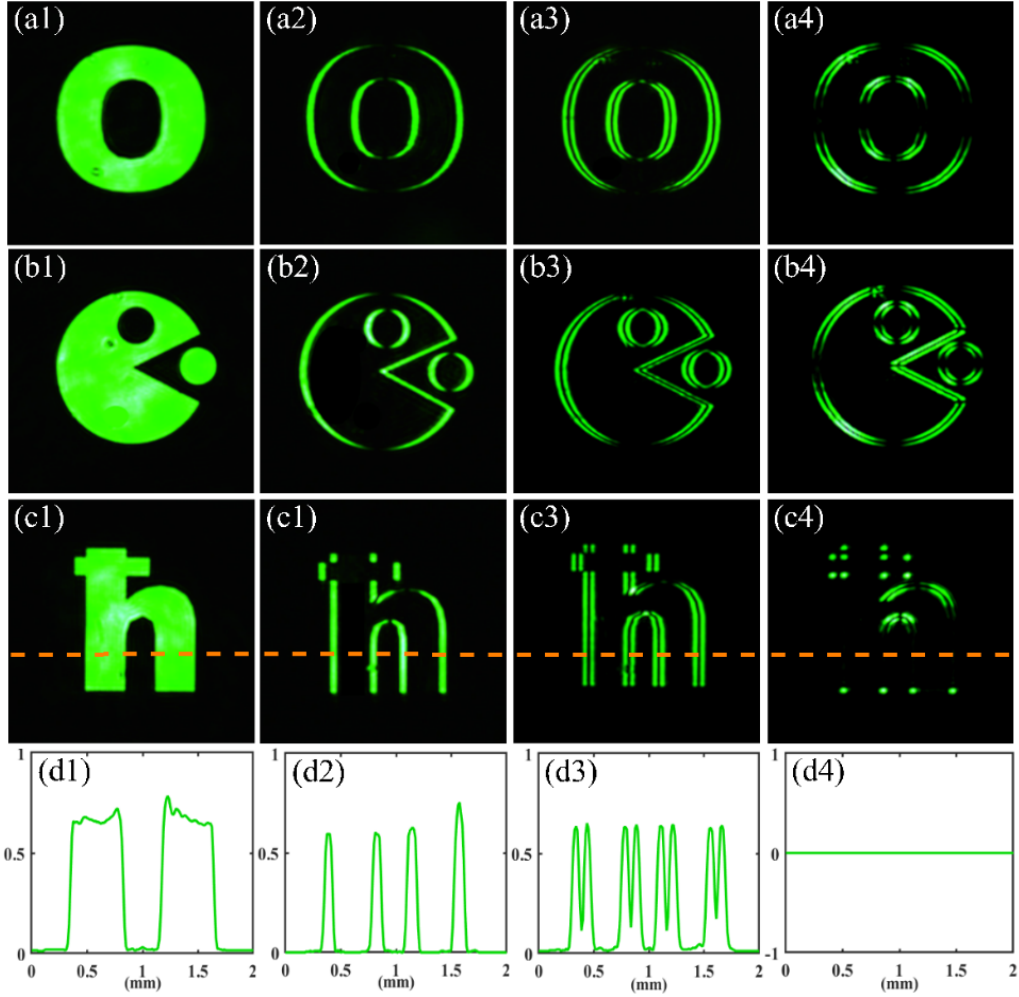

**FIG. 3.** All-optical spatial higher-order differentiator based on cascaded operation for amplitude objects. (a1)-(c1) show the input intensity images. (a2)-(c2), (a3)-(c3), and (a4-(c4) are obtained by performing $\partial E_{in}/\partial x$, $\partial^2 E_{in}/\partial x^2$ and $\partial^2 E_{in}/\partial x \partial y$ operations on the input image, respectively. The intensity curve data for (d1)-(d4) are taken from (c1)-(c4) at the orange dashed line.

After passing the second differentiation unit, the double edge effect is generated by the second-order spatial differentiation $\partial^2 E_{in}/\partial x^2$ and $\partial^2 E_{in}/\partial x \partial y$ as shown in Figs. 4(a3)-4(c4). The second-order spatial differentiation can present more accurate feature recognition on the image and find the phase mutation point of the first-order differentiation to identify the edges. In addition to performing second-order differentiation based on the metasurfaces, the cross-polarization component of the dipole scattering field can be written in the form of second-order spatial differentiation, which arises from the transverse nature of the electromagnetic field. Introducing this intrinsic differential operation into quantum microscopy can provide a possible path to the field of nondestructive biomedical imaging.

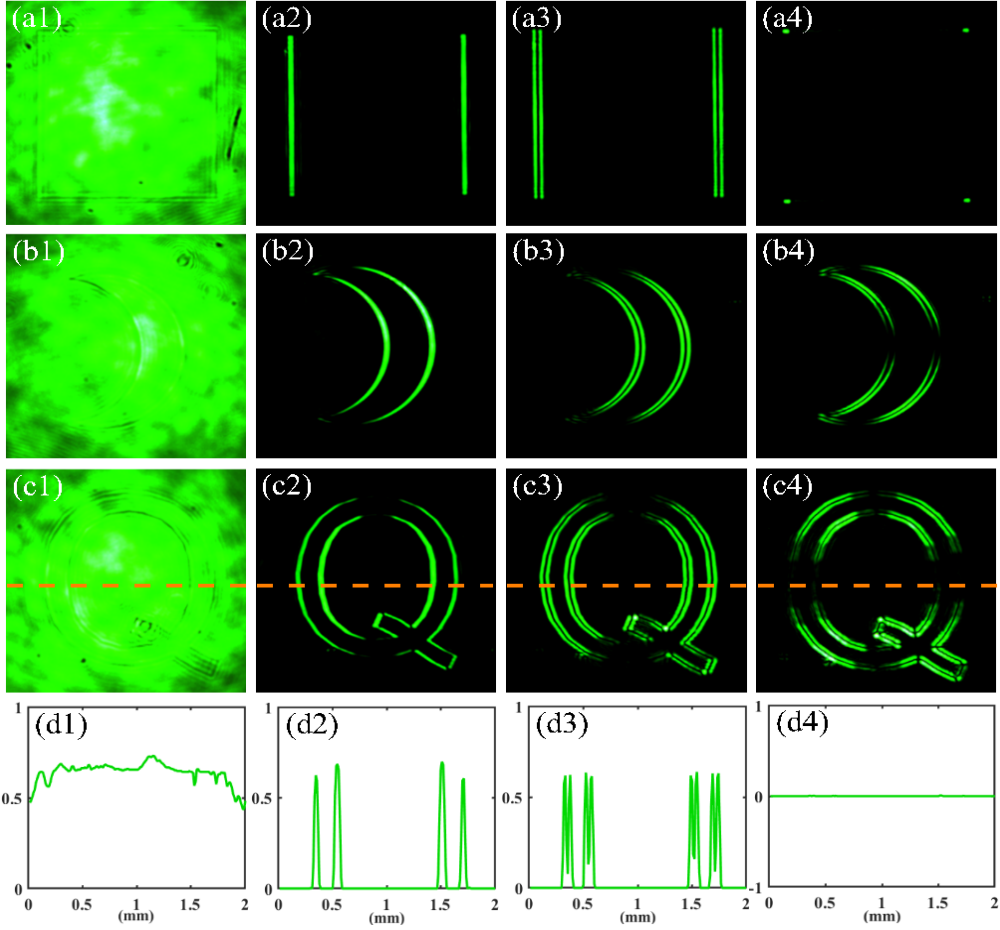

**FIG. 4.** All-optical spatial higher-order differentiator based on cascaded operation for phase objects. (a1)-(c1) are the bright field image without placing the metasurface. (a2)-(c2), (a3)-(c3), and (a4)-(c4) are obtained by performing $\partial E_{in}/\partial x$, $\partial^2 E_{in}/\partial x^2$ and $\partial^2 E_{in}/\partial x \partial y$ operations on the bright-field image, respectively. (d1)-(d4) are the intensity cross sections of (c1)-(c4).

# CONCLUSIONS

In conclusion, we propose a scheme to implement an all-optical higher-order spatial differentiator based on cascaded operations. The second-order spatial differentiator can be realized by cascading two first-order differential operation units. The all-optical image edge detection of intensity and phase objects through this cascaded system is experimentally demonstrated. Furthermore, the cascaded scheme is not limited to differential units, but can also synthesize other analog computing units. This is an important step toward the development of advanced optical analog computing networks and providing a possible way for multifunctional all-optical image processing. In practical applications, although there is transmission loss in each unit, the transmission efficiency of the cascaded system can still be enhanced by photon recycling operations.

# ACKNOWLEDGMENTS

This work was supported by the National Natural Science Foundation of China (12174097) and Natural Science Foundation of Hunan Province (2021JJ10008).